\documentclass[12pt]{article}
\usepackage{graphicx,epsfig,psfig,color,latexsym}
\newcommand{\bc}{\begin{center}}
\newcommand{\ec}{\end{center}}
\newcommand{\be}{\begin{equation}}
\newcommand{\ee}{\end{equation}}
\newcommand{\bea}{\begin{eqnarray}}
\newcommand{\eea}{\end{eqnarray}}
\newcommand{\ba}{\begin{array}}
\newcommand{\ea}{\end{array}}
\newcommand{\lb}{\label}
\newcommand{\rf}{\ref}
\newcommand{\bfg}{\begin{figure}[htbp]}
\newcommand{\efg}{\end{figure}}
\newcommand{\bbt}{\bibitem}
\newcommand{\pr}{Phys. Rev. }
\newcommand{\np}{Nucl. Phys. }
\newcommand{\prl}{Phys. Rev. Lett. }

\newcommand{\ap}{Ann. Phys. (N.Y.) }
\newcommand{\pl}{Phys. Lett. }

\newcommand{\nc}{Nuovo Cimento }
\newcommand{\ptp}{Prog. Theor. Phys. }

\newcommand{\epj}{Eur. Phys. J. }
\newcommand{\cmp}{Comm. Math. Phys. }

\begin{document}

\begin{flushright}
IPNO-DR-04-09
\end{flushright}
\vspace{0.5 cm}
\bc
{\large \textbf{Relevance of the strange quark sector \protect \\
in chiral perturbation theory\footnote{Talk given at the Conference 
Quark Confinement and the Hadron Spectrum VI, Villasimius, Cagliari,
Italy, 21-25 September 2004.}}}
\vspace{1. cm}

H. Sazdjian\\
\vspace{0.25 cm}
\textit{Groupe de Physique Th\'eorique,
Institut de Physique Nucl\'eaire\footnote{Unit\'e Mixte de Recherche
8608.},\\
Universit\'e Paris XI, F-91406 Orsay Cedex, France\\
\footnotesize{E-mail: sazdjian@ipno.in2p3.fr}}
\ec
\par
\renewcommand{\thefootnote}{\fnsymbol{footnote}}
\vspace{0.75 cm}

\bc
{\large Abstract}
\ec
\par
Results obtained in recent years in the strange quark sector of  
chiral perturbation theory are reviewed and the theoretical relevance 
of this sector for probing the phase structure of QCD at zero 
temperature with respect to the variation of the number of massless
quarks is emphasized.
\par   
\vspace{0.5 cm}
\par
\newpage

\section{Chiral perturbation theory} \lb{s1}
Chiral perturbation theory (ChPT) is an effective theory of QCD
at low energies, in which the dynamical degrees of freedom are those
of the pseudoscalar Goldstone bosons ($\pi$, $K$, $\eta$) of the 
chiral group $SU(3)\times SU(3)$. That theory was developed by
Weinberg \cite{w1} and applied in more detail to the QCD case by 
Gasser and Leutwyler \cite{gl1,gl2}. The effective lagrangian 
$\mathcal{L}_{eff}$ is the most general locally chiral invariant
lagrangian in the presence of external source terms. It contains 
an infinite series of terms constructed out of the pseudoscalar 
meson fields and the external sources. The series is arranged 
according to an increasing power of derivatives and quark masses. 
At low energies, the Goldstone boson interactions are weak
(they are of second order in the derivatives and of first
order in the quark masses) and therefore a perturbative 
calculation of their transition amplitudes is meaningful.
The expansion parameter is essentially the order of magnitude
of the external momenta or of the quark masses divided by the 
hadronic mass scale (which is of the order of 1 GeV). 
\par
The counting rules of the dimensionalities of various diagrams 
are based on the counting of the numbers of external momenta, of 
the mass terms  and of loops, each quark mass being equivalent to 
a momentum to the power 2. Furthermore, because of the weakness of 
the interactions at the tree level at low momenta, each loop 
introduces two additional powers of momenta, thus contributing to 
nonleading orders \cite{w1}. Generally, there are terms of order 
$O(p^2)$, then terms of order $O(p^4)$, $O(p^6)$, etc. The effective
lagrangian takes the following corresponding expansion:
\be \lb{e1}
\mathcal{L}_{eff}=\mathcal{L}_2+\mathcal{L}_4+\mathcal{L}_6
+\ldots \ ,
\ee
each index indicating the power of the momenta that will emerge
from a tree-level calculation of a process.
It is evident that at low energies it is the terms with smallest
indices that will be the dominant ones.
\par
The effective lagrangian is renormalizable order by order in
perturbation theory. The term $\mathcal{L}_2$ contributes,
through its one-loop effects, to the $O(p^4)$ terms, which are
generally ultra-violet divergent. Those divergences are then
absorbed in redefinitions of the coupling constants contained
in the terms of the lagrangian $\mathcal{L}_4$. The latter,
in turn, produces, with $\mathcal{L}_2$, one-loop divergences of 
order $O(p^6)$ that are absorbed, together with the two-loop 
divergences of the lagrangian $\mathcal{L}_2$, by the coupling 
constants of the lagrangian $\mathcal{L}_6$, and so forth.
\par
At each order of the perturbation series there are a certain 
number of coupling constants, called low energy constants (LEC),
that are order parameters of spontaneous chiral symmetry breaking.
They are expected to be determined from experimental
data. In addition, one also encounters the quark masses 
$m_u$, $m_d$, $m_s$.
\par
At order $O(p^2)$, one has two LECs: $F_0$, which is the pion
weak decay constant $F_{\pi}$ in the chiral limit, and
$B_0=-<0|\overline uu|0>_0/F_0^2$, which is proportional to 
the quark condensate in the vacuum in the chiral limit \cite{gl2,gmor}.
At order $O(p^4)$, one has 10 (observable) LECs, called 
$L_i$, $i=1,\ldots,10$ \cite{gl2}. At order $O(p^6)$, one has 90
(observable) LECs, called $C_i$, $i=1,\ldots,90$ \cite{bce}.
It is worthwhile to emphasize that not all LECs enter in a
definite process.
\par        
The theory relatively simplifies if one sticks to processes
related to the nonstarnge sector of the quarks ($u$, $d$).
The chiral group now becomes $SU(2)\times SU(2)$. Here, the 
strange quark can be considered as heavy and the corresponding 
field integrated out.
\par
The $SU(2)\times SU(2)$ version of ChPT contains less LECs 
than its $SU(3)\times SU(3)$ version. At order $O(p^2)$ one still
has two LECs, $F$ and $B$, the analogs of $F_0$ and $B_0$, but
now considered in the $SU(2)\times SU(2)$ chiral limit. At
order $O(p^4)$, there are 7 LECs, called $\ell_i$, $i=1,\ldots,7$
\cite{gl1}. At order $O(p^6)$, there are 53 LECs, called
$c_i$, $i=1,\ldots,53$ \cite{bce}. 
\par
A detailed study of the elastic $\pi\pi$ scattering amplitude up
to order $O(p^6)$ was done by several groups \cite{bcegs,kmsf,gkms,
cgl1}. The rate of convergence of ChPT seems rather satisfactory:
$O(p^4)$ effects represent approximately 20-25\% of the global
quantities under consideration, while $O(p^6)$ effects represent
7-8\% of the contributions. Recent experimental data from the 
E865 experiment at Brookhaven about $K_{e4}$ decay \cite{p}, which 
provides information about the $\pi\pi$ scattering amplitude near 
threshold through the final state interaction, have confirmed the 
hypothesis that the quark condensate parameter $B$ is the leading 
order parameter of chiral symmetry breaking \cite{gmor,cgl1,cgl2}. 
This means that the QCD vacuum is very similar to a ferromagnetic 
medium, as far as chiral symmetry breaking is concerned. 
\par 
Extension of ChPT to $SU(3)\times SU(3)$ allows the study of
sectors involving the $K$ and $\eta$ mesons. But here, the strange
quark mass $m_s$ is not as small as the non-strange ones, $m_u$
and $m_d$. From the tree-level relation \cite{gmor}
$2m_s/(m_u+m_d)=2m_K^2/m_{\pi}^2-1\simeq 25$ one guesses that
ChPT might converge more slowly than in the $SU(2)\times SU(2)$
case. Apart from that aspect, which by itself leads to unavoidable
complications, it was emphasized by Descotes, Girlanda and Stern 
\cite{dgs} that the type of dependence of physical quantities on 
the strange quark mass $m_s$ might reveal some important theoretical
features of QCD.
\par

\section{Phase transition in the number of massless quarks} \lb{s2}
It is known that $SU(N_c)$ gauge theories might undergo a zero
temperature chiral phase transition when the number $N_f$ of massless 
fermions (in the fundamental representation) reaches some critical 
value $N_f^*<11N_c/2$ \cite{bz}. The argument goes as follows. For 
$N_f<11N_c/2$, one has an 
asymptotically free theory, while for $N_f>11N_c/2$ asymptotic
freedom is lost. From the first two terms of the beta-function
one infers the existence of an infra-red stable fixed point appearing
when $N_f$ reaches from below a critical value $N_{f0}(<11N_c/2)$.
In perturbation theory, one has $N_{f0}\simeq 34N_c/13$. When the 
value of $N_f$ further increases and reaches the vicinity of
$11N_c/2$ from below, then the domain of variation of the
effective coupling constant becomes tiny and the theory becomes
fully perturbative, in the infra-red and in the ultra-violet, 
reducing to a conformal theory. Such theories, because of the
smallness of the coupling constant, do not undergo spontaneous
chiral symmetry breaking, neither display confinement \cite{atw}. 
On the other hand, at small values of $N_f$, one has chiral symmetry
breaking and confinement \cite{th,cw,fsby,cg}. Therefore, there
should exist a critical value of $N_f$, $N_f^*$, such that
$N_{f0}<N_f^*<11N_c/2$, where the theory undergoes a chiral
phase transition (Fig. \rf{f1}).   
\par
\bfg
\vspace*{0.5 cm}
\bc
\begin{picture}(0,0)%
\epsfig{file=f1.pstex}%
\end{picture}%
\setlength{\unitlength}{2960sp}%
\begingroup\makeatletter\ifx\SetFigFont\undefined%
\gdef\SetFigFont#1#2#3#4#5{%
  \reset@font\fontsize{#1}{#2pt}%
  \fontfamily{#3}\fontseries{#4}\fontshape{#5}%
  \selectfont}%
\fi\endgroup%
\begin{picture}(6719,1078)(2379,-4117)
\put(8851,-3961){\makebox(0,0)[lb]{\smash{\SetFigFont{14}{16.8}{\familydefault}{\mddefault}{\updefault}{\color[rgb]{1,0,0}$N_f$}%
}}}
\put(6076,-4036){\makebox(0,0)[lb]{\smash{\SetFigFont{14}{16.8}{\familydefault}{\mddefault}{\updefault}{\color[rgb]{1,0,0}$N_f^*$}%
}}}
\put(5101,-4036){\makebox(0,0)[lb]{\smash{\SetFigFont{14}{16.8}{\familydefault}{\mddefault}{\updefault}{\color[rgb]{1,0,0}$N_{f0}$}%
}}}
\put(3451,-4036){\makebox(0,0)[lb]{\smash{\SetFigFont{14}{16.8}{\familydefault}{\mddefault}{\updefault}{\color[rgb]{1,0,0}$3$}%
}}}
\put(2851,-4036){\makebox(0,0)[lb]{\smash{\SetFigFont{14}{16.8}{\familydefault}{\mddefault}{\updefault}{\color[rgb]{1,0,0}$2$}%
}}}
\put(7201,-4036){\makebox(0,0)[lb]{\smash{\SetFigFont{14}{16.8}{\familydefault}{\mddefault}{\updefault}{\color[rgb]{1,0,0}$11N_c/2$}%
}}}
\end{picture}

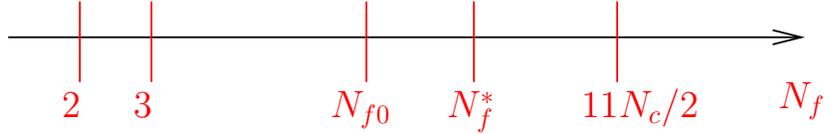
\caption{Domains of $N_f$. At $N_{f0}$, reached from below, an 
infra-red fixed point appears. At $N_f^*$ a chiral phase transition 
from the Goldstone mode ($N_f<N_f^*$) to the Wigner mode 
($N_f^*<N_f<11N_c/2$) occurs. Above $11N_c/2$ asymptotic freedom is 
lost.}
\lb{f1}
\ec
\efg
It is evident that at $N_f^*$ all chiral order parameters 
should have vanished. Among those, $F_0$ would be the last one 
to vanish, since it is the fundamental order parameter of chiral 
symmetry breaking, directly related to the Goldstone theorem.
\par
The precise value of $N_f^*$ is dependent on the dynamical models
that are used to evaluate it. Appelquist, Terning and Wijewardhana 
\cite{atw}, combining perturbation theory and gap equation calculations
find $N_f^*\simeq 4N_c=12$. Lattice calculations give a rather wide
range of values. Kogut and Sinclair \cite{ks}, Brown \textit{et al.}
\cite{brw} find $8\leq N_f^*\leq 12$; Iwasaki \textit{et al.} \cite{i} 
find $N_f^*\simeq 6$, while Mawhinney \cite{mw} finds $N_f^*\simeq 4$. 
Within the instanton liquid model, Velkovsky and Shuryak \cite{vs}
find $N_f^*\simeq 6$.  
\par
From another viewpoint, Descotes, Girlanda and Stern have studied
the dependence on $N_f$ of various chiral order parameters \cite{dgs}.
Using properties of the Dirac operator in the background gluon field
in euclidean space (placed in a box) and bounds derived by Vafa and
Witten \cite{vw} concerning its eigenvalues, they obtain the
following inequalities when the number of massless fermions changes
from $N_f$ to $N_f+1$:
\bea 
\lb{e2}
F_0^2[N_f+1] &<& F_0^2[N_f],\\
\lb{e3}
|<\overline uu>|_{(N_f+1)} &<& |<\overline uu>|_{N_f}.
\eea
These inequalities, which do not hinge on any hypothesis about the
existence of a chiral phase transition, are manifestly compatible,
at least locally, with the behaviors of order parameters as expected
from such an hypothesis.
\par
In summary, the eventual existence of a chiral phase transition in
$N_f$ would have the tendancy to decrease the values of order parameters
with increasing $N_f$. The slope of the variation would strongly depend
on the value of the critical point $N_f^*$. It would be stronger
for smaller $N_f^*$. In the real world one does not have much freedom
to vary $N_f$. The only possibility that we have is to vary $N_f$
from 2 ($SU(2)\times SU(2)$) to 3 ($SU(3)\times SU(3)$). (For
$N_f=1$, chiral symmetry is destroyed by the $U_A(1)$-anomaly.)
Possibly small values of $N_f^*$ (4-6, say) would induce rather
strong variations of order parameters in passing from $N_f=2$ to
$N_f=3$. Therefore, phenomenological studies of such effects
would represent indirect tests about the vicinity of $N_f^*$.
In particular, quantities that are Zweig-rule suppressed in the
large-$N_c$ limit (almost true for $N_f=2$ and $N_c=3$), should be 
enhanced for $N_f=3$ \cite{dgs}. Those mainly concern scalar meson 
sectors and the LECs $L_4$ and $L_6$. Simultaneously, loops of the 
strange quark might provide important contributions and destabilize 
certain results obtained in the large-$N_c$ limit \cite{dgs}.
\par

\section{Phenomenology} \lb{s3}
The above problem was first studied by Moussallam \cite{m1}.
He calculated the ratio of the quark condensate evaluated in
a theory with three massless quarks to the condensate evaluated
in a theory with two massless quarks:
\be \lb{e4}
R_{32}\equiv \frac{<\overline uu>_{N_f=3}}{<\overline uu>_{N_f=2}}.
\ee
In the $N_f=2$ case, the mass of the strange quark is fixed at
its ``physical'' value; but since the latter is still small compared
to the massive hadron masses, one can use perturbation theory for it
and keep only the leading contribution in $m_s$. One thus obtains, at
the one-loop level:
\be \lb{e5}
R_{32}=1-\frac{m_sB_0}{F_{\pi}^2}\Big[32L_6(\mu)-
\frac{1}{16\pi^2}\Big(\frac{11}{9}\ln(\frac{m_sB_0}{\mu^2})+
\frac{2}{9}\ln(\frac{4}{3})\Big)\Big]+O(m_s^2),
\ee
where $\mu$ is the renormalization mass. 
The quantity $m_sB_0$ can be replaced by its tree-level expression,
$(m_K^2-m_{\pi}^2/2)$. $L_6$ is then evaluated from the correlator
of scalar-isoscalar densities, $(\overline uu+\overline dd)$ and 
$\overline ss$:
\be \lb{e6}
\int d^4x e^{\displaystyle ip.x}\langle 
T[(\overline uu(x)+\overline dd(x))\overline ss(0)]\rangle_c.
\ee
This is precisely a Zweig-rule violating term. It is evaluated by
saturating the intermediate states with $\pi\pi$ and $K\overline K$
states, yielding the pion and kaon scalar form factors. One obtains
coupled Muskelishvili--Omn\`es equations. Use of experimental values
of phase shifts and phases leads to:
\bea
\lb{e7}
& &L_6(m_{\eta})=(0.6\pm 0.2)\times 10^{-3},\\
\lb{e8}
& &R_{32}=0.46\pm 0.27.
\eea
The last result indicates a strong variation of the quark condensate
when passing from two massless quarks to three.
\par
A similar study, by a different method, was also done in Ref. \cite{ds},
confirming the above conclusions. $O(p^6)$ effects, estimated by means 
of a resonance model and the sigma-model, do not seem to qualitatively 
change the above results \cite{m2}.
\par
An important quantity in the strange quark sector is the $\pi K$
elastic scattering amplitude. The tree-level (current algebra),
$O(p^2)$, values of the $S$-wave isospin 1/2 and 3/2 scattering 
lengths had been calculated by Weinberg \cite{w2}:
\be \lb{e9}
a_0^{1/2}=0.14,\ \ \ \ \ \ a_0^{3/2}=-0.07.
\ee
(In units of $m_{\pi}^{-1}$.)
\par
The scattering amplitude at the one-loop level, $O(p^4)$, was 
calculated by Bernard, Kaiser and Meissner \cite{bkm}. The 
scattering lengths become:
\be \lb{e10} 
a_0^{1/2}=0.19,\ \ \ \ \ \ a_0^{3/2}=-0.05.
\ee
\par
Until recently, experimental knowledge of the scattering lengths
was very poor. As for $\pi\pi$, it is not possible to realize
direct scattering experiments at low energies, because pions and 
kaons decay. One must then use extrapolations of high energy data 
to low energies. Recently, a detailed evaluation of the low-energy
$\pi K$ elastic scattering amplitude was done by B\"uttiker,
Descotes-Genon and Moussallam \cite{bdm}, by means of Roy
and Steiner type equations \cite{r,s}. Those equations use 
dispersion relations, crossing symmetry, unitarity and 
partial-wave analysis, together with high-energy data, to
reconstruct the elastic scattering amplitude at low energies. 
The method was already used for $\pi\pi$ scattering 
\cite{blgn,bfp,pp,fp,cgl1}. In $\pi K$, one ends up with six 
coupled integral equations. Solutions with rather small 
uncertainties have been obtained for the scattering lengths
\cite{bdm}:
\be \lb{e11}
a_0^{1/2}=0.224\pm 0.022,\ \ \ \ \ \ a_0^{3/2}=-0.045\pm 0.008.
\ee
\par
$O(p^6)$ effects in $\pi\pi$ and $\pi K$ scattering were 
evaluated (for $N_f=3$) by Bijnens, Dhonte and Talavera \cite{bdt,b}.
The corresponding LECs are calculated by resonance saturation methods. 
They do an overall fit to all existing data (scattering amplitudes, 
form factors, masses, etc.), leaving the Zweig-rule violating LECs
$L_4$ and $L_6$ as free parameters. They obtain several sets of
results, depending on which experimental quantities optimization is 
imposed by varying slightly the resonance parameters. For the set 
producing the best fit with the scattering amplitudes, they find for
the $\pi K$ scattering lengths:
\be \lb{e12}
a_0^{1/2}=0.220,\ \ \ \ \ \ a_0^{3/2}=-0.047.
\ee
The relevant LECs are:
\be \lb{e13}
L_4=0.2\times 10^{-3},\ \ \ \ \ \ \ L_6=0.0\times 10^{-3}.
\ee
The above values of the scattering lengths match, within the 
allowed uncertainties, those obtained from the Roy--Steiner 
extrapolation method of high energy data [Eq. (\rf{e11})]. The 
values of the LECs are also compatible with a small violation of 
the Zweig rule. At this point one might conclude that ChPT is
rapidly converging, without sizable Zweig-rule violating effects.
However, the overall fit of Refs. \cite{bdt,b} displays in some
instances contradictory aspects. One notices that in other sectors
(mainly the scalar ones) $O(p^6)$ effects are more important than
$O(p^4)$ effects, indicating bad convergence (in particular in the
pionic sector, which, in the $SU(2)\times SU(2)$ case had a rapid 
convergence). In such cases, the meaning of the $O(p^4)$ LECs
$L_4$ and $L_6$ becomes questionable. Perhaps optimization with 
respect to global convergence of $SU(3)\times SU(3)$ should be tried.
\par
In this respect, Descotes-Genon, Fuchs, Girlanda and Stern have 
proposed a different method of evaluation of high-order effects
\cite{dfgs}. They suggest to isolate those terms which might be
sensitive to Zweig-rule violating effects (four in all) and to
treat them nonperturbatively, while treating the rest perturbatively.
The method was already applied to the $\pi\pi$ scattering case;
its application to the other sectors could still reduce the
existing uncertainties. 
\par
Finally, we mention here some future useful experiments about
the $\pi K$ system.
\par
The process
\be \lb{e14}
D\longrightarrow \pi K e\nu,
\ee
could be analyzed in the FOCUS experiment at FermiLab. It would
give information, through the final state interaction,
about the elastic $\pi K$ phase shifts. It plays an analogous role
as the $K_{\ell 4}$ decay for $\pi\pi$ scattering.
\par
The process
\be \lb{e15}
\tau\longrightarrow \pi K\nu,
\ee
could be analyzed in the CLEOIII experiment at Cornell.
It would give information about the $K_{\ell 3}$ form factors.
\par
The observation and measurement of the properties of the hadronic
atom $(K^+\pi^-)_{at.}$ would also give complementary informations
about the scattering lengths. Hadronic atoms are Coulomb bound states 
of charged hadrons, which generally decay under the effect of the
strong interactions into neutral isospin partners. Thus the above 
atom would mainly decay as
\be \lb{e16}
(K^+\pi^-)_{at.}\longrightarrow K^0\pi^0.
\ee
The lifetime of the atom depends essentially on the combination
$(a_0^{1/2}-a_0^{3/2})$ of the $\pi K$ scattering lengths, while
the energy level splittings depend upon the combination
$(2a_0^{1/2}+a_0^{3/2})$ \cite{dgbt,up,t,bnnt}. The experimental
study of the pionium (the $\pi^+\pi^-$-atom) is currently done at 
CERN in the DIRAC experiment. For the $K^+\pi^-$-atom, projects
are being prepared. To precisely reconstruct the strong interaction
scattering lengths from the hadronic atom properties, one needs
to take into account isospin breaking and electromagnetic radiative
corrections, as well as relativistic corrections. A recent theoretical
study of the $K\pi$-atom was done by Schweizer \cite{sw}.    
\par
In conclusion, the study of the strange quark sector up to order
$O(p^6)$ offers the possibility of a full test of $SU(3)\times SU(3)$
ChPT and at the same time of an indirect probe of a possible phase 
transition in the number of massless flavors in QCD.
\par
\vspace{0.25 cm}
\noindent
\textbf{Acknowledgements}: This work was supported in part by the 
European Community network EURIDICE under contract No. 
HPRN-CT-2002-00311
\par

\end{document}